\documentclass[doublecol]{epl2} 
\pdfoutput=1
\usepackage{cite}
\usepackage{numprint}
\npthousandsep{,}\npproductsign{*}\npdecimalsign{.}
\usepackage{listings}
\usepackage{caption}
\usepackage{floatrow}
\usepackage{float}
\usepackage{enumitem}
\newcounter{nalg} 
\renewcommand{\thenalg}{\arabic{nalg}} 
\DeclareCaptionLabelFormat{algocaption}{Algorithm \thenalg} 

\lstnewenvironment{algorithm}[1][] 
{    
    \refstepcounter{nalg}
    \lstset{ 
        frame=tB,
        numbers=left, 
        numberstyle=\tiny,
        basicstyle=\scriptsize, 
        keywordstyle=\color{black}\bfseries\em,
        keywords={,input, output, foreach, while, begin, end, if, else, for, return, } 
        numbers=left,
        xleftmargin=.04\textwidth,
    	abovecaptionskip=0.7em,
    	aboveskip=2em,
        captionpos=b,
        #1 
    	}
	\captionsetup{labelformat=algocaption,labelsep=colon}
}
{}

\title{Discovering novel ingredient pairings in molecular gastronomy\\ using network analysis}
\shorttitle{Molecular Gastronomy \& Network Analysis} 

\author{Aleksander Klju\v cev\v sek \and Luka Krapi\'{c}}
\shortauthor{A. Klju\v cev\v sek, L. Krapi\'{c}}

\institute{University of Ljubljana, Faculty of Computer and Information Science, Ljubljana, Slovenia}

\abstract{ Molecular gastronomy is a distinct sub-discipline of food science that takes an active role in examining chemical and physical properties of ingredients and as such lends itself to more scientific approaches to finding novel ingredient pairings. With thousands of ingredients and molecules, which participate in the creation of each ingredient's flavour, it can be difficult to find compatible flavours in an efficient manner. Existing literature is focused mainly on analysis of already established cuisine based on the flavour profile of its ingredients, but fails to consider the potential in finding flavour compatibility for use in creation of completely new recipes. Expressing relationships between ingredients and their molecular structure as a bipartite network opens up this problem to effective analysis with methods from network science. We describe a series of experiments on a database of food using network analysis, which produce a set of compatible ingredients that can be used in creation of new recipes. We expect this approach and its results to dramatically simplify the creation of new recipes with previously unseen and fresh combinations of ingredients.}

\begin{document}

\maketitle 

\section{\label{sec_intro}Introduction}

Essential part of any great dish is harnessing compatible flavours from its ingredients. This includes, but is not limited to, the knowledge of which spices, herbs and other flavourings accentuate particular ingredients best. The tried and tested method of trial and error has been the go to method for finding such compatibility throughout the history. While this approach produced numerous timeless combinations\textemdash such as basil with tomatoes and mozzarella cheese or apples with cinnamon\textemdash and over time resulted in classic cuisines, its one flaw is that it depends on the imagination of the chef and his willingness to try as many random combinations as possible. There are many possibilities that make for good combinations, but would probably never get tried due to how different they seem\textemdash white chocolate with caviar, oysters with passion fruit, etc. 

Today, in a modern kitchen, it is possible to utilize the scientific method in finding compatible flavours without relying on a single person's taste. Once we discovered enough about volatile compounds of ingredients, it became clear that flavour compatibility is based on molecular similarity of different ingredients \cite{Ouweland1997, Blumenthal2008}. A single ingredient can be composed of hundreds, sometimes even more than a thousand flavour compounds. This makes it difficult to efficiently analyse and compare large numbers of ingredients and limits the creative landscape of chefs looking to invent new recipes. One approach that proved successful in analysis and comparison of ingredients in network analysis is construction of a flavour network \cite{Ahn2011}. With this approach ingredients and volatile compounds are presented as nodes in a graph, while edges connect ingredients with compounds they contain. We use this approach as basis for our work. We expand on it by looking for groups of complimentary ingredients by joining nodes that have similar molecular profiles into clusters \cite{TheMetabolomicsInnovationCentre2015}. Furthermore, we use partially labelled data in conjunction with semi supervised community detection \cite{Zhang2013} to account for nuances that can't be described by shared molecular profile alone, to improve the accuracy of our algorithm and the method used in Ahn's approach \cite{Ahn2011}. 

Since this field of food science is relatively unexplored, data on compatible flavour parings are scarce and limited. With our approach we show that finding complimentary pairings can be relatively easy to implement and execute on any food database containing enough data. 

\section{Related work}

One of the pioneers in finding novel ingredient pairings was H. Blumenthal, who popularized molecular gastronomy through his restaurant and recipe books  \cite{Blumenthal2008}. Ahn et al.\cite{Ahn2011} uses the flavour network to compare and contrast Western and Eastern cuisine and to see whether we prefer to use ingredients that share more flavour compounds. Their findings confirm their hypothesis, but only for Western cuisine, while Eastern cuisine displays preference for ingredients with different flavour base. An extension of this technique is further used to analyse specific local cuisines. Bogojeska et al. \cite{Bogojeska2016} analyse Macedonian cuisine and its flavours, while Jain et al. \cite{Jain2015} seek to determine whether cuisines of different regions of India follow the Eastern cuisine pattern discovered by Ahn et al.. Caporaso et al.\cite{caporaso2015true} look closely at how using different types of vegetable oils affects chemical properties and volatile profile of a traditional Neapolitan pizza. Interactions between food and drugs has been well documented, but Jovanovik et al. \cite{Jovanovik2015} show how different drug categories have different distributions of negative effects in different parts of the world due to differences in regional cuisines. In a more narrow analysis, Ruiz et al. \cite{Ruiz2015} look at molecular constitution as one of the methods in creating novel chocolates. Using Ahn's approach \cite{Ahn2011}, Tackx et al. \cite{Tackx2015} generalize it in trying to develop new metrics for studying intricate patterns observed in real networks, which standard metrics do not account for.

\section{Data}

\begin{figure}[ht]
\caption{Degree distribution of projected graph\textemdash ingredient network.}\label{initialGraphDegreeDistribution}
\centering\includegraphics[width=8cm,height=6cm]{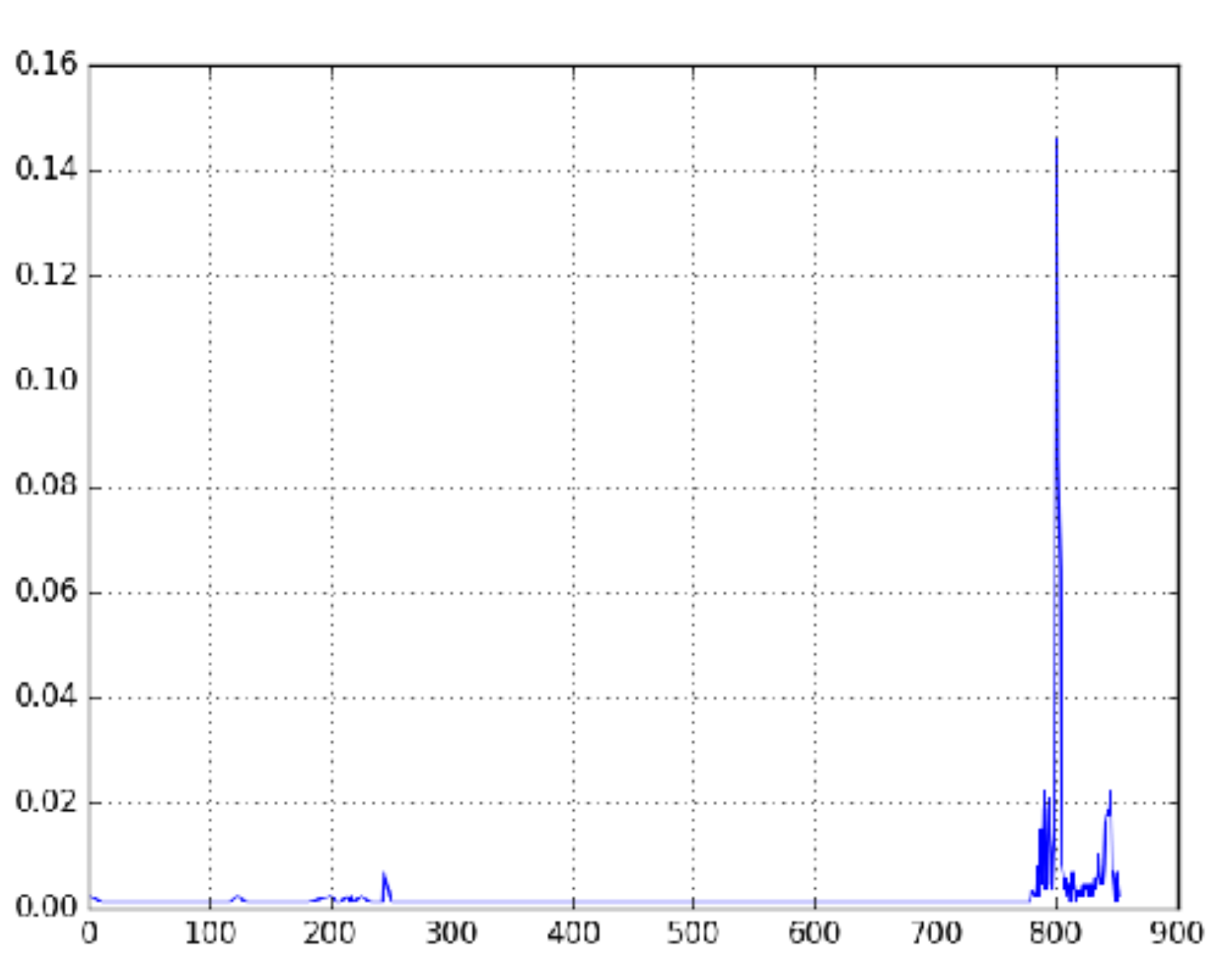}
\end{figure}

All data about food and food components comes from FooDB \cite{TheMetabolomicsInnovationCentre2015}. The ingredients and compounds from this database form two disjoint sets of a bipartite graph\textemdash a type of graph with two disjoint sets of nodes inside which no two nodes share a link\textemdash while edges link ingredients with compounds if the former contains the latter. Such representation allows us to efficiently express information about similarities of ingredients based on the number of shared components. The graph consists of \numprint{16269} nodes, with \numprint{856} of these nodes representing ingredients, while the rest represent the flavour molecules and \numprint{106113} edges. Degree distribution of this graph projected into ingredient space is represented in Figure \ref{initialGraphDegreeDistribution}. Most ingredients share at least one component, with average degree $<k>=\numprint{767,533}$ making the graph very dense $(density is \numprint{0,898})$ and in need of filtering, which is described in the next section.

\begin{figure}[ht]
\caption{Degree distribution of ingredient network after filtration.}\label{initialGraphDegreeDistribution-filtered}
\centering\includegraphics[width=8cm,height=6cm]{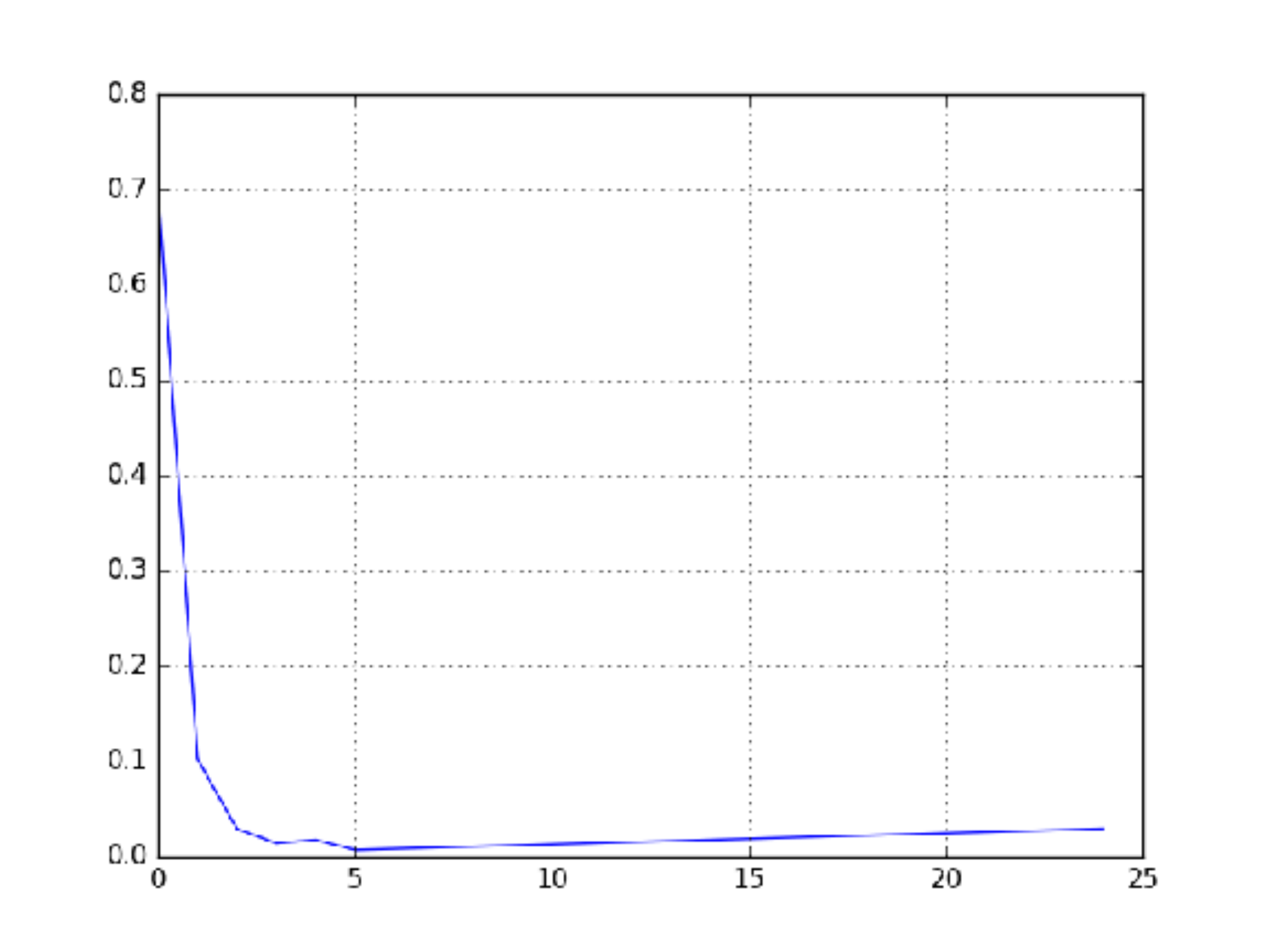}
\end{figure}

In order to verify our findings, we use information about known compatible ingredient pairings in Western and Eastern cuisine. Data for Western cuisine is extracted from one of the most popular recipe websites Epicurious \cite{CondeNast2016}, while data for Eastern cuisine is extracted from Rasa Malaysia \cite{Corp.2016} in order to avoid Western interpretation of Eastern cuisine. We collect \numprint{1000} highly rated recipes (rating 4/4) from Epicurious, which are chosen at random from all the different types of recipes, e.g., appetizers, lunches, dinners, sides, desserts, etc., and \numprint{507} recipes from Rasa Malaysia, also from various categories. Ingredients are then parsed and matched to ingredients in our database. Parsed recipes for Western cuisine contain on average \numprint{7.62} distinct ingredients, while recipes for Eastern cuisine contain on average \numprint{10,77} unique ingredients. In order to make both sets easier to compare, we reduce the size of the larger one to the size of the smaller one, so that they contain \numprint{507} recipes each.

These data form the body of knowledge used for training and testing our classification algorithm.

\section{Methods}

In order to get a more compact representation of our original bipartite graph, we project it into ingredient space. This projection produces a simple weighted graph where weights of edges correspond to number of shared components between two adjacent nodes. Resulting flavour network is essentially similar to the one constructed by Ahn et al. \cite{Ahn2011}. 

Our initial projected graph consists of \numprint{856} nodes and \numprint{328504} links, and is dense to the point of a complete graph. Consequently some filtration to reduce this is necessary. The filtration process is done as follows: we define similarity of two nodes based on the weight of the link they share. Furthermore, we assume every node is similar to at least one of its neighbours. Filtration is done locally for every node. Filtering with factor $p \in [0, 1]$ means removing all links below value $p\ast w_{max}$ for a given node, where $w_{max}$ is maximum link weight, i.e., the largest similarity it shares with any other ingredient.

Separate training, validation, and testing datasets for semi-supervised community detection are built from Western and Eastern recipes.
We take standard measures of $80\% - 10\% - 10\%$ for training, validation and testing  respectively, equally distributed among both sets of recipes. Each dataset contains unique samples of recipes. Based on Ahn's work \cite{Ahn2011}, we assume each pair of ingredients from one of Western recipe forms a good flavour pairing, and each pair of ingredients from one of Eastern recipe forms bad flavour pairing within the context of Western cuisine. Testing and validation datasets are split into two pairs: Western testing set, Western validation set, Eastern testing set and Eastern validation set. Each such set is represented as a list of ingredients which belong to one recipe. Training dataset is also split in two pairs: Western and Eastern training set. Each such set is represented as a simple weighted graph in which nodes are ingredients from projected flavour network and edge between nodes exist if adjacent nodes occur together in at least one common recipe from its own training dataset. Weight of the edges correspond to number of recipes adjacent edges share in their own datasets. This ensures that ingredients paired more often share a stronger connection.

To ensure greater reliability of recipe training data, we take ingredient pairs which occur in more recipes as more probable, containing more reliable information. Both graphs are then filtered with the same procedure described earlier for flavour network. Since we are trying to establish solid pairings within a certain context\textemdash Western or Eastern cuisine\textemdash we treat pairs that occur in both datasets as discrepancies and omit them.

Let C be a connected component from Western graph which contains such discrepancy. Then we say node $(n_1 \in V(C)$ has discrepancy if there exists edge $(n_1 n_2)$ in Eastern graph where $(n_2 \in V(C))$. Let $(n \in V(C))$ be a node with discrepancy. Let $(D_n = {nu \vert u \in V(C) \wedge nu \in E(Eastern\_graph)})$ and $(D_p = {nu \vert u \in V(C) \wedge nu \in E(C)})$. Furthermore let  $(d_n = \sum_{e \in D_n} weight(e))$ and $(d_p = \sum_{e \in D_p}  weight(e))$. Then value of discrepancy for node n is is \(d_n = \max\{\frac{d_p}{d_n}, \frac{d_n}{d_p}\}\). Procedure for removing discrepancies in training data is explained below.

\begin{algorithm}[caption={Sanity check}]
  input: Western_graph, Eastern_graph
  while there are discrepancies in the data:
    ccs = connected components(Western_graph)
    for component in ccs:
      n = node with maximal discrepancy d
      if d < 3:
        remove edges Dp from Western_graph
        remove edges Dn from Eastern_graph
      else if dp > dn:
        remove edges Dn from Eastern_graph
      else
        remove edges Dp from component
\end{algorithm}

Since removing edges from component can split it into several components, we first iterate over all connected components and then calculate connected components again. We set threshold value for discarding both \(D_p\) and \(D_n\) edges to 3 because we don't expect same ingredient pairs to appear in both recipe sets in the similar ratio.

Next, we employ improved semi supervised algorithm for community detection \cite{Zhang2013} to detect communities of good flavour pairings. In order to be able to use this algorithm, we must assume communities are disjunctive and that our projected graph has transitive property, i.e., every ingredient pair inside one community is a compatible pairing. Zhang et al. \cite{Zhang2013} compare multiple internal algorithms for community detection, of which we choose Infomap due to its good performance. We use the implementation of Infomap that can be found on \cite{Mapequation.org2015}. Infomap's starting parameters were set to:
\begin{itemize}[noitemsep]
\setlength\itemsep{0.4em}
\item undirected graph
\item optimized for two level partition of the network
\item five repetitions before solution is found
\end{itemize}

To evaluate performance of our model under given parameters we calculate sensitivity and specificity on validation sets. Sensitivity shows how well our model can find good ingredient pairings for Western kitchen and is calculated as the average of sensitivity values for every recipe in Western recipe list. Sensitivity for a single recipe is calculated as a number of ingredient pairs in the same community divided by a number of every combination of two food components in a given recipe. Specificity shows how well our model can find good ingredient pairings for Eastern kitchen and is calculated as the average of specificity values for every recipe in the Western recipe list. Specificity for a single recipe is calculated as a number of ingredient pairs which are in different community divided by a number of every combination of two food components in a given recipe.

\begin{algorithm}[caption={Implementation overview}, label={alg1}]
input: raw flavour network G, 
       Western recipe list W,
       Eastern recipe list E,
       filter parameter Ff,
       filter parameter Fr,
       knowledge percent p.

output: model, accuracy

H = project graph and filter (G, Ff)
Wf = construct graph and filter (W, Fr)
Ef = construct graph and filter (E, Fr)
Wg, Eg = Sanity check (Wf, Ef)

Wt = construct validation list (W)
Et = construct validation list (E)

G = community detection(H, Wg, Eg, p)

A = accuracy (G, Wt, Et)

return G, A
\end{algorithm}

We iterate over parameters $F_{f}$ (filtration parameter for projected flavour network), $F_{r}$ (filtration parameter for training graphs), and $p$ of the algorithm to find best possible values with which to train model. Final\textemdash best\textemdash parameters are selected based on the sensitivity and specificity. Iteration range of said parameters is $[0,1]$ with step $\Delta = \numprint{0,05}$. Evaluation of parameters is repeated five times, with training and validation sets chosen anew for each of the five repetitions. Final score is the average of results from each of the five repetitions. This is done to more precisely represent how good a model is under given parameters.

To see how well our algorithm generalizes, we train the model with best parameters and evaluate it on the testing set. We repeat this procedure 10 times and report average scores.

\section{Results}

We obtain close to \numprint{300000} unique pairings, sorted on their compatibility. Examples of good pairings as well as bad ones can be seen in Tables \ref{tab-compatible} and \ref{tab-incompatible}.

\begin{table}[]
\centering
\caption{Five of the most compatible ingredient pairings}\label{tab-compatible}
\begin{tabular}{ll}
Ingredient A & Ingredient B \\
\hline
sweet orange & lemon        \\
pear         & apple        \\
apple        & plum         \\
parsley      & sweet orange \\
apricots     & broccoli    \\
\hline
\end{tabular}
\end{table}

\begin{table}[]
\centering
\caption{Five of the least compatible ingredient pairings}\label{tab-incompatible}
\begin{tabular}{ll}
Ingredient A & Ingredient B  \\
\hline
avocado      & mushrooms     \\
elderberry   & cherry tomato \\
nougat       & brassicas     \\
blackcurrant & onion         \\
apple        & crustaceans  \\
\hline
\end{tabular}
\end{table}

Since establishing good ingredient pairings is as important as establishing bad ones, we both sensitivity and specificity to find the best model and the best ingredient pairings. 

Best model is obtained for parameters $F_f = \numprint{1,0}$ and $F_r = \numprint{0,15}$, which results in $sensitivity=\numprint{0,5}$  and $specificity=\numprint{0,8}$. Applied on testing sets, this model results in $sensitivity=\numprint{0,56}$ and $specificity=\numprint{0,8}$. 

\begin{figure}[h]
\caption{Results for the model with parameters $F_f = \numprint{1,0}$ and $F_r = \numprint{0,15}$. Blue line represents sensitivity, red line specificity.}\label{best-model-image}
\centering\includegraphics[width=8cm,height=6cm]{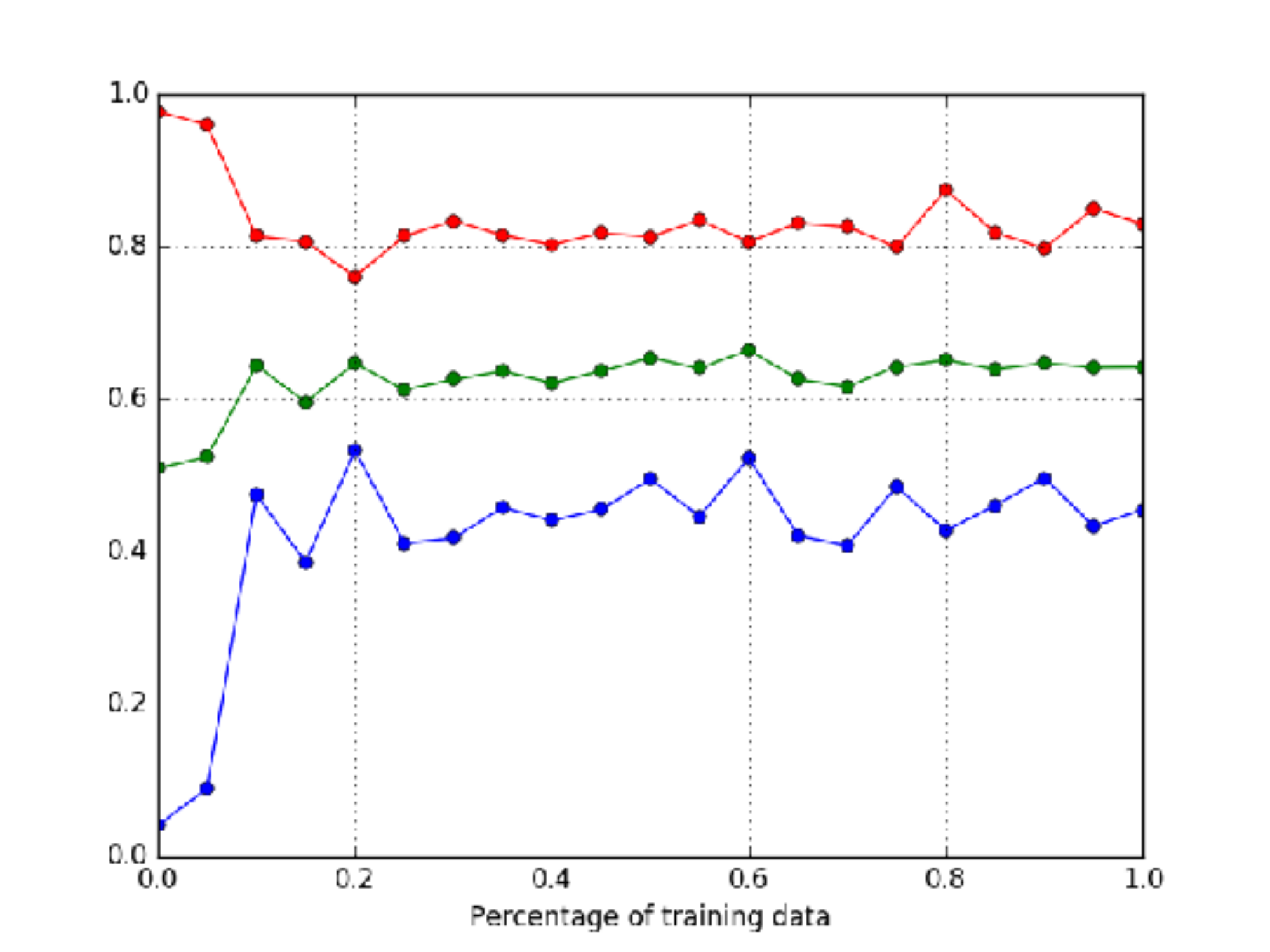}
\end{figure}

Adding only $\numprint{10}\%$ of our training data, which amounts to \numprint{40} recipes for each, Western and Eastern recipes, our model becomes effective in classifying ingredient pairings. We can get models with arbitrary high individual accuracies, but they cannot be considered good classifiers.

For every ingredient pairing in Tables \ref{tab-compatible} and \ref{tab-incompatible} we cross-checked result with classification according to our best model. Results are shown in Tables \ref{tab-class-compatible} and \ref{tab-class-incompatible}.

\begin{table}[h]
\centering
\caption{Classification results for compatibility of ingredients from Table \ref{tab-compatible} in the context of Western cuisine. Reverse holds for the Eastern cuisine.}\label{tab-class-compatible}
\begin{tabular}{llc}
Ingredient A & Ingredient B & compatibility \\
\hline
sweet orange & lemon        & no \\
pear         & apple        & yes \\
apple        & plum         & yes \\
parsley      & sweet orange & no \\
apricots     & broccoli    & yes \\
\hline
\end{tabular}
\end{table}

\begin{table}[h]
\centering
\caption{Classification results for compatibility of ingredients from Table \ref{tab-incompatible} in the context of Western cuisine. Reverse holds for the Eastern cuisine.}\label{tab-class-incompatible}
\begin{tabular}{llc}
Ingredient A & Ingredient B & Compatibility \\
\hline
avocado      & mushrooms     & yes \\
elderberry   & cherry tomato & no \\
nougat       & brassicas     & no \\
blackcurrant & onion         & no \\
apple        & crustaceans  & no \\
\hline
\end{tabular}
\end{table}

\section{Discussion}

Methodology devised in this paper shows we can use several network analysis techniques in finding novel ingredient pairings. Using already established pairings, it is then possible to refine these results even further. 

While validating close to \numprint{300000} ingredient pairings would be practically impossible, we show that a portion of results can be verified using already established well known pairings. Main reason for choosing a subset to work with is the lack of availability of recipes including all the ingredients in the food database. Since food database contains as many distinct ingredients as possible, finding tested, well rated recipes which combine all of them is practically impossible\textemdash especially within the limited contexts of a specific cuisine. 

Another improvement lies in the interpretation of Ahn et al. \cite{Ahn2011} findings; if Western cuisine is based on pairing together ingredients with as similar molecular structure as possible, then using Eastern cuisine recipes as negative examples helps us produce compatible pairings for that criteria with greater accuracy. In order to improve results in the context of Eastern kitchen, we can reverse the sets and use Western recipe set as negative examples and Eastern set as positive examples. Lower sensitivity results narrow down this interpretation for Eastern cuisine to pairing together ingredients whose molecular structure does overlap, but contrary to Western cuisine's approach where the overlap is maximized, Eastern ingredient pairs overlap only partially, resulting in a broader flavour profile.

\section{Conclusion and future work}

Our methodology proves itself as a useful heuristic in approach to classification of compatible ingredient pairings. Using it to analyse food pairings within specific contexts of Western and Eastern cuisine serves to further clarify Ahn's findings \cite{Ahn2011}. While Eastern cuisine does tend to use less compatible ingredients than its Western counterpart, it does so in pursuit of a broader palate. In the end, there still has to be some compatibility between the ingredients to tie the whole dish together.

We devised our methodology to be as generally applicable as possible, which is especially important given that better results can be derived as we get more and better data. One shortcoming that could easily be accounted for with more expansive data is that ingredients can develop different flavours depending on their age\textemdash bananas, for example, develop more sugar and consequently higher glycemic index as they ripen while losing some of the vitamins and minerals\textemdash while seared meat develops a recognizable taste through Maillard reaction \cite{vega2013the}. Another possible improvement could account for substitutions, where one ingredient is substituted not for the taste, but in order to get desired consistency\textemdash substituting cake flour with combination of regular flour and corn starch, resulting in flour with lower levels of protein content, which makes for less chewy texture\textemdash or some other property, e.g., color. 

\bibliographystyle{plain}

\end{document}